# Robust method to provide exponential convergence of model parameters solving LTI plant identification problem

**Anton Glushchenko, Vladislav Petrov, Konstantin Lastochkin**

Stary Oskol technological institute n.a. A.A. Ugarov (branch) NUST "MISIS"

a.glushchenko@sf-misis.ru

**Abstract.** The scope of this research is a problem of parameters identification of a linear time-invariant (LTI) plant, which 1) input signal is not frequency-rich, 2) is subjected to initial conditions and external disturbances. The memory regressor extension (MRE) scheme, in which a specially derived differential equation is used as a filter, is applied to solve the above-stated problem. Such a filter allows us to obtain a limited regressor value, for which a condition of the initial excitation (IE) is met. Using the MRE scheme, the recursive least-squares (RLS) method with the forgetting factor is used to derive an adaptation law. The following properties have been proved for the proposed approach. If the IE condition is met, then: 1) the parameter error of identification is a limited value and converges to zero exponentially (if there are no external disturbances) or to a bounded set (in the case of them) with an adjustable rate, 2) the parameters adaptation rate is a finite value. The above-mentioned properties are mathematically proved and demonstrated via simulation experiments.

## Introduction

The development of adaptive control systems for real process units (e.g. heating, electromechanical, chemical) of various branches of industry and complex objects (e.g. airplanes) is an actual problem since most of them are quasi-stationary [1, 2], i.e. have parametric uncertainty. For such plants, it is required to provide high control quality: 1) for a wide range of setpoint values (it is often a set of constant values), 2) under the condition of disturbances of different nature and noise.

Adaptive systems can be divided into direct and indirect ones [1, 2]. As for the indirect systems, it is necessary to identify the plant, which model is usually introduced as a regression dependence of type $y = \theta^{\mathrm{T}}\varphi$. Its parameters $\theta$ are adjusted with the help of gradient methods or recursive least-squares (RLS) online according to the real-time values of the regressor $\varphi$ and the plant output $y$. As far as direct methods are concerned, the same linear regression is often used as a control law, which parameters are adjusted online using the above-mentioned methods. The Lyapunov second method and Ultimate Uniform Boundedness are most often used for stability analysis of such systems.

Despite a great deal of research in this field, the adaptive control systems have not become widespread in the industry yet, as there are fundamental problems with their application. It is a choice of adaptation rate [3, 4], the requirement of the persistent excitation of the regressor [5], drift of the adjusted parameters of the mentioned regression because of the noise, disturbances, and limitations on values of a control action signal [6].

The main among them is the restrictive condition of the persistent excitation (PE) of the regressor to provide parameter convergence [5]. Only if the PE condition is met, it is proved [5] that the model parameters will converge to the ideal values (which are considered to be constant) exponentially. This will guarantee the robustness of the control system. Otherwise, the regression parameters will not become equal to their ideal values and/or they drift when t→∞. But, in reality, it is difficult to meet the PE condition. The setpoint is a constant (or set of constants) for many technological processes. The PE condition is not satisfied in this case. Well-known projection operator and $e$- and σ-modifications [2] do not solve the problem under consideration. They guarantee that all signals are bounded in the case of unmodeled dynamics and/or noise, but they do not provide convergence of the regression parameters to their ideal values.

Therefore, as far as adaptive control and identification are concerned, nowadays one of the main lines of research is to relax the PE condition. This is confirmed, in particular, by the paper [5] with a detailed review of known approaches to solve this problem. To begin with the application of filters [7, 8] for the regressor extension, and composite adaptation [9], when both tracking and parameter errors are used in the adaptation law.

Before we move on to methods to relax PE condition, the basic principles of filtering will be discussed. Regressor filtration gives the opportunity to implement the following principle: the higher the adaptation rate, the higher the parameters convergence rate (and, as a result, the approximation quality). If the filtration is not used, then there is an optimal value of the adaptation rate. If we exceed it, then we face the convergence rate decrease. A filtration is also a tool for the regressor extension, for which two main approaches should be mentioned: The Dynamic Regressor Extension (DRE) [10] and the Memory Regressor Extension (MRE) [8, 11]. In the case of the DRE, many different filters $H(p)$ (or signal delay blocks) are applied to the regressor vector $\varphi$ and the function $y$. This allows obtaining as many regression equations as the number of the unknown regression parameters $\theta$. At the same time, it is necessary to somehow find parameters values for each filter, and the identification algorithm will start functioning only when the last filter (delay) output becomes a non-zero value.

In the case of the MRE, the left and right sides of the regression equation are multiplied by the regressor vector $\varphi$. Then only one filter $H(p)$ is applied to the multiplications $\varphi\varphi^T$ and $y\varphi^T$. Comparing to the DRE, it makes the problem of the filter parameters choice less complicated (as the number of such parameters is much lower). The gradient method is then applied to the extended regressor to estimate the regression parameters. But their ideal values (these are values, which will reduce the parameter error to zero) can be found only in the case if the PE condition is met.

Considering composite adaptation methods, the regression adaptation law contains an additional term of sum, which is used to indicate the current control quality. This should improve such quality, but this fact is not proved [5]. The main term (responsible for the parameter convergence) is usually derived using the above-mentioned methods.

All considered methods provide exponential parameter convergence only under the PE condition. Let solutions to relax the PE condition be considered.

One of such methods, which is based on the DRE, is DREM [10], in which a matrix regressor is transformed into a scalar one by additional multiplication of the filtered regression (according to the DRE principles) by a matrix, which is adjoint to one of the filtered regressor. This operation is time-consuming, but the scalar regressor and a separate equation for each regression parameter significantly simplify further calculations. This scheme also provides exponential convergence of parameters only in the case of the PE. But if the regressor does not belong to $L_2$, then the asymptotic convergence is guaranteed [12]. From the practical point of view, it is not quite clear what gives the relaxation to "not in $L_2$" regressor from the PE one, and what types of setpoint for real technological processes meet this condition.

In most practical cases, the regressor is excited only within a limited time interval (for example, within an initial time range for a constant setpoint) – the initial excitation (IE) [13]. If exponential convergence of regression parameters is provided in this case, it would significantly increase the chances of practical implementation of adaptive identification and control schemes. The basic approach, based on the MRE and used to relax the PE to IE, is concurrent learning [14]. Considering this method, the adaptation law depends not only on the current value of the product of the regressor vector by the error, but also on some previous values of this product saved in a data stack. This guarantees the parameter convergence without the PE, but having the IE, because, even after "attenuation" of the current regressor, its stored values will allow providing the exponential convergence. The drawbacks of the method are: 1) the question of what data to store in the data stack, and 2) not studied question of the correctness of the given adaptation law under conditions of non-zero initial conditions and external disturbances.

The method proposed in [13] is based on a similar idea. But, in this case, the previous values of the regressor and the error are taken into consideration not with the help of the data stack, but because of the fact that, using the MRE method to extend the regressor, the filter in the form of an integrator is applied instead of an aperiodic link-based filter. By analogy with [14], it also allows providing exponential convergence of regression parameters in the case of the IE. In [5] it is asserted that such a scheme has the following disadvantages: open-loop integration of a positive semi-definite matrix, the drift of the regression parameters under a condition of external disturbances, and the unstudied question of such scheme effectiveness in the case of non-zero initial conditions.

This paper is to focus on the method described in [13]. First of all, its improvement will be proposed allowing functioning under the condition of disturbances (to avoid the drift of the regression parameters). For this purpose, it is proposed to use a specially derived differential equation instead of a pure integrator as a filter. Then the formulas to identify the plant, using the RLS with the exponential forgetting factor, will be derived: 1) to obtain a regression parameters adaptation law, which will be different from that given in [13]; 2) to obtain the equation to adjust the convergence rate. Further, the properties of such a solution will be studied. It will be shown that, when the proposed scheme is used and the IE condition is satisfied, the parameter error of identification is a limited value and exponentially converges to zero (when there are no disturbances) or to a bounded set (in the case of perturbations) with adjustable convergence rate.

## Problem statement

Consider a problem of parameters identification of a continuous-time LTI plant given by (1) [2].

$$\begin{cases} \dot{x}(t) = Ax(t) + Bu(t), \ x(0) = x_0, \\ y(t) = C^T x(t) + w(t), \end{cases} \quad (1)$$

where $x \in R^n$ is a state vector, $x_0 \in R^n$ is a vector of the plant states initial condition, $u \in R$ is a control action, $y \in R$ is a measurable plant output signal, $w \in R$ is a bounded disturbance ($w(t) \leq w_{max}$); $A \in R^{n \times n}$ is a stable state matrix, $B \in R^{n \times 1}$ is an input matrix, $C \in R^{n \times 1}$ is an output matrix. The values of $A$, $B$, and $C$ elements are unknown. It is supposed that only $u$ and $y$ signals are measurable. The plant (1) can also be represented as an input-output model (2).

$$y(t) = \frac{Z(s)}{R(s)} u(t) + \frac{C^T \text{adj}(sI - A)}{R(s)} x_0 + w(t), \quad (2)$$

where $Z(s) = \sum_{i=0}^{m} b_i s^i$, $R(s) = \sum_{i=0}^{n} a_i s^i$ are characteristic polynomials of the numerator and denominator ($n \geq m$, $a_n = 1$) respectively. Values of parameters $a_i$ and $b_i$ of the characteristic polynomials $Z(s)$ and $R(s)$ are to be found as a result of the identification.

The input-output model (2) can be parametrized [2, 15] as a regression model (3) under the condition of filtration of $u(t)$ and $y(t)$ signals using stable $n^{\text{th}}$-order filter $\Lambda$.

$$y(t) = \theta^T \omega(t) + \eta_0(t) + w(t),$$

$$\theta^T = \begin{bmatrix} b_m \\ \vdots \\ b_0 \\ a_{n-1} - \lambda_{n-1} \\ \vdots \\ a_0 - \lambda_0 \end{bmatrix}^T ; \; \omega(t) = \begin{bmatrix} \dfrac{\alpha_m(s)}{\Lambda(s)} u(t) \\ -\dfrac{\alpha_{n-1}(s)}{\Lambda(s)} y(t) \end{bmatrix} ; \; \alpha_i(s) = \begin{bmatrix} s^i \\ s^{i-1} \\ \vdots \\ 1 \end{bmatrix}, \quad (3)$$

where $\Lambda(s) = s^n + \sum_{i=0}^{n-1} \lambda_i s^i$ is a Hurwitz polynomial, $\omega(t) \in R^{n+m+1}$ is a known regressor vector, $\theta \in R^{n+m+1}$ is a vector of ideal constant parameters of the regression model (3), $\eta_0$ is an exponentially decaying function caused by the initial condition influence on (1). The function $\eta_0(t)$ can be described using a function generator of type (4).

$$\begin{cases} \dot{v}(t) = \Lambda_c v(t), \; v(0) = B_0 x_0, \\ \eta_0(t) = C_0^T v(t), \end{cases} \quad (4)$$

where $v \in R^n$, $\Lambda_c$ is a stable matrix corresponding to the characteristic polynomial $\Lambda(s)$, the matrices $C_0 \in R^n$ and $B_0 \in R^{n \times n}$ are chosen to model the generator (4) according to the following condition: $C^T_0 \{adj(sI - \Lambda_c)\} B_0 = C^T \{adj(sI - A)\}$. Then $\eta_0(t)$ can be written as (5).

$$\eta_0(t) = C_0^T e^{\Lambda_c t} B_0 x_0 \quad (5)$$

A regression (6) with adjustable parameters $\hat{\theta}(t)$ is introduced to find the unknown vector $\theta$.

$$\hat{y}(t) = \hat{\theta}^T(t) \omega(t) \quad (6)$$

The parameter adjustment law is usually obtained using the first-order optimization methods or Lyapunov second method [2]. If $\eta_0(t) = 0$ and $w(t) = 0$, then the adaptation law can be written as (7).

$$\dot{\hat{\theta}}(t) = -\Gamma \omega [\hat{y}(t) - y(t)]^T = -\Gamma \omega [\hat{\theta}^T(t)\omega(t) - \theta^T \omega(t)]^T = -\Gamma \omega(t) \omega^T(t) \tilde{\theta}(t), \quad (7)$$

where $\tilde{\theta}(t) = \hat{\theta}(t) - \theta$ is a parameter error between (3) and (6), $\Gamma$ is a matrix of adaptation rate. As parameters $\theta$ are constant, so parameter error function $\tilde{\theta}(t)$ can be written as a solution (8) of the differential equation (7).

$$\tilde{\theta}(t) = e^{-\Gamma \int_0^t \omega(\tau) \omega^T(\tau) d\tau} \tilde{\theta}(0); \quad (8)$$

It is well-known [2, 16] that, considering (8), the parameter error $\tilde{\theta}(t)$ converges to zero exponentially only if the PE condition (9) is satisfied.

*Definition 1*: The regressor $\omega$ is persistently excited ($\omega \in$ PE) if $\forall t \geq 0 \; \exists T > 0$ and $\alpha > 0$ such that

$$\int_{t}^{t+T} \omega(\tau)\omega^{T}(\tau)d\tau \geq \alpha I, \tag{9}$$

where *I* is an identity matrix, and α is an excitation degree.

For an instance, in the case $n+m+1 = 2$, the regressor, which meets the PE condition (9), is a vector $\omega = [1; \sin(t)]^T$.

It has been proved in [17] that $\omega \in$ PE if and only if the control action $u(t)$ is frequency-rich.

*Definition 2*: The control action $u(t)$ is frequency-rich if, using the Fourier transform, it could be written as a finite series (10), so as $N \geq n+m+1$ and $\forall k \neq j$, $\varphi_k \neq \varphi_j$, $A_k \neq 0$.

$$u(t) = \sum_{k=1}^{N} A_k \sin(\varphi_k t) \tag{10}$$

The control action $u(t)$ is formed by the control system for the plant (1). In many cases, the control objective does not require it to be frequency-rich. As a result, the real control action signal could not be written as (10) as it is needed to identify the plant. So the requirement $\omega \in$ PE to provide the convergence of $\tilde{\theta}(t)$ to zero has the following well-known limitations: 1) it is hard to be satisfied in most practical tasks; 2) it is impossible to check whether the criterion (9) is satisfied in real time, as it depends on future values of the regressor.

Another interpretation of the persistent excitation condition (9-10) is the need for all elements of the vector ω to be non-zero to provide convergence of $\tilde{\theta}(t)$ to zero. Following the setpoint schedule, real plants (1) function both in dynamic modes, when all elements of vector ω are not equal to zero, and in static ones, when elements of vector ω, corresponding to estimations of $y^1,\ldots, y^{n-1}$ and $u^1,\ldots, u^m$, are equal to zero. In this case, we can say that the regressor $\omega \notin$ PE, but it is excited at the initial interval of time (during the transient until the static mode is reached).

*Definition 3*: Considering $t \in [0;\infty)$, the regressor $\omega(t)$ is excited during the initial time interval $[0;T]$ ($\omega \in$ IE), if there exist $T>0$ and $\alpha > 0$ such that:

$$\int_{0}^{T} \omega(\tau)\omega^{T}(\tau)d\tau \geq \alpha I, \tag{11}$$

where *I* is an identity matrix, α is an excitation degree.

For example, if $n+m+1 = 2$, then the regressor, for which the initial excitation condition is met, is a vector $\omega = [1; e^{-t}]^T$.

The initial excitation means that the regressor has energy within a certain initial time interval. The IE condition is much less restrictive than the PE one, and in contrast: 1) can be checked online to be satisfied; 2) is provided not by special conditions (10) on external signals, but by the natural behavior of the dynamic plants in the course of transients. In [13, 14] a criterion has been obtained, which allows checking in real-time whether the initial excitation condition (11) is satisfied or not.

*Definition 4*: The regressor $\omega(t)$ satisfies the initial excitation condition within time interval $[0; T]$ if an increasing time sequence $S = \{t_i\}|_{i=1}^{p}$ exists, where $t_p \leq T$ and $t_1 > 0$, and $rank(W)=n+m+1$, where *W* is defined as (12).

$$W = \begin{bmatrix} \omega(t_1) & \omega(t_2) & \cdots & \omega(t_p) \end{bmatrix} \tag{12}$$

*Remark 1*: Based on definitions (11) and (12), it is obvious that the condition of initial excitation can be checked to be met by calculation of either matrix (12) rank or determinant of the integral (11).

The initial excitation condition is less restrictive than the persistent excitation one. In addition, according to *Definition* 4 and *Remark* 1, it can be checked online. So, in practice, there is a need for the adaptation law to estimate parameters θ, which provides exponential convergence to zero of the parameter error $\tilde{\theta}(t)$ when ω∈IE, but not the PE. Therefore, taking into account the shortcomings of the existing solutions [13, 14] mentioned in the introduction section, the aim of this research is to develop a law to estimate the unknown parameters θ of the linear regression (3) with the following properties:

– when the initial excitation condition (ω∈IE) is met, the parameter error $\tilde{\theta}(t)$ converges to zero exponentially;

– the rate of convergence can be made arbitrarily high if the value of a certain parameter in the adaptation scheme is increased;

– the adaptation law is robust with respect to the functions $\eta_0(t)$ and $w(t)$.

## Main results

First of all, the regressor extension method is applied to equation (3) [8, 11] to obtain (13) to develop an adaptation law, which has the properties specified in the problem statement section.

$$y(t)\omega^T(t) = \theta^T \omega(t)\omega^T(t) + \eta_0(t)\omega^T(t) + w(t)\omega^T(t) \quad (13)$$

Both the left and right sides of the equation (13) are multiplied with the function $e^{-\beta t}$, where $\beta>0$ is a memory factor, to obtain (14).

$$e^{-\beta t} y(t)\omega^T(t) = e^{-\beta t}\theta^T \omega(t)\omega^T(t) + \underbrace{e^{-\beta t}\left[\eta_0(t) + w(t)\right]\omega^T(t)}_{\dot{\varepsilon}(t)} \quad (14)$$

The filter (15) is applied to the extended regressor $e^{-\beta t}\omega(t)\omega^T(t)$ and new output $e^{-\beta t} y(t)\omega^T(t)$, taking into account initial condition and disturbance $\dot{\varepsilon}(t)$.

$$\begin{cases} \dot{\Omega}(t) = e^{-\beta t}\omega(t)\omega^T(t) \\ \dot{\Upsilon}(t) + \dot{\varepsilon}(t) = e^{-\beta t} y(t)\omega^T(t) + e^{-\beta t}\left[\eta_0(t) + w(t)\right]\omega^T(t) \end{cases} \quad (15)$$

*Remark 2*: Filters (15) are similar to the ones proposed in [13], which have been used for the synthesis of the integral component of the PI law of the plant parameters estimation. However, unlike them, the proposed filters (15) allow obtaining a bounded regressor $\Omega(t)$ and the function $\Upsilon(t)$.

Given the filtration (15), equation (14) can be written as (16).

$$\dot{\Upsilon}(t) = \theta^T \dot{\Omega}(t) + \dot{\varepsilon}(t) \quad (16)$$

The left and right sides of (16) are integrated with respect to time to obtain (17).

$$\Upsilon(t) = \theta^T \Omega(t) + \underbrace{\int_0^t e^{-\beta\tau}\left[\eta_0(\tau) + w(\tau)\right]\omega^T(\tau)\,d\tau}_{\varepsilon(t)} \quad (17)$$

The new regressor $\Omega(t)$ and output $\Upsilon(t)$ can be found from (15) as (18).

$$\begin{cases} \Omega(t) = \int_0^t e^{-\beta\tau}\omega(\tau)\omega^T(\tau)\,d\tau \\ \Upsilon(t) = \int_0^t e^{-\beta\tau} y(\tau)\omega^T(\tau)\,d\tau \end{cases} \quad (18)$$

The regressor $\Omega(t)$ has the following important properties: 1) as it follows from the first equation of (18), $\Omega(t)$ is a positive semidefinite function, such that $\Omega(t) > 0\ \forall t \geq 0$; 2) $\Omega(t)$ is a time function, which increases to its finite limit. To prove the second property of the regressor $\Omega(t)$, its upper bound (19) is obtained using the mean value theorem.

$$\Omega(t) = \int_0^t e^{-\beta\tau}\omega(\tau)\omega^T(\tau)\,d\tau \leq \delta^2 \int_0^t e^{-\beta\tau} I\,d\tau = \frac{\delta^2}{\beta}\left(1 - e^{-\beta t}\right) I \leq \frac{\delta^2}{\beta} I$$
$$\delta = \sup_{t \geq 0}\max_{1 \leq i \leq n+m+1}|\omega_i| \quad (19)$$

It is important to show for further development of the adaptation law that the disturbance $\varepsilon(t)$ is bounded. Taking into account that $\omega(t) \in L\infty$ and $w(t) \in L\infty$, and $\eta_0(t)$ is the exponentially damped function (5), this fact ($\varepsilon(t) \in L\infty$) is proved (20) using the mean value theorem.

$$\varepsilon(t) = \int_0^t e^{-\beta\tau}\left[\eta_0(\tau) + w(\tau)\right]\omega^T(\tau)\,d\tau = \int_0^t e^{-\beta\tau} C_0^T e^{\Lambda_c \tau} B_0 x_0 \omega^T(\tau)\,d\tau +$$
$$+ \int_0^t e^{-\beta\tau} w(\tau)\omega^T(\tau)\,d\tau \leq \delta C_0^T B_0 x_0 (\beta - \Lambda_c)^{-1}\left(1 - e^{(\Lambda_c - \beta)t}\right) + \delta w_{max}\beta^{-1}\left(1 - e^{-\beta t}\right) \leq \quad (20)$$
$$\leq \underbrace{\left(\delta C_0^T B_0 x_0 \left\|(\beta - \Lambda_c)^{-1}\right\| + \delta w_{max}\beta^{-1}\right)}_{\varepsilon_{max}}$$

*Remark 3*: The fundamental role in the disturbance $\varepsilon(t)$ boundedness is played by the multiplication of (14) by the function $e^{-\beta t}$, which is shown in (15).

Similar to the equation (6), the regression with adjustable parameters $\hat{\theta}(t)$ is introduced for (17). Then the parameter error function between the regression and function (17) is written as (21).

$$\hat{\Upsilon}(t) - \Upsilon(t) = \underbrace{\hat{\Upsilon}(t) - \Upsilon(t)}_{\tilde{\Upsilon}(t)} = \hat{\theta}^T(t)\Omega(t) - \theta^T\Omega(t) - \varepsilon(t) = \tilde{\theta}^T(t)\Omega(t) - \varepsilon(t) \quad (21)$$

Now, following the recursive least squares method to develop the identification loop for $\hat{\theta}(t)$ at time moment $t$, let the measurements $\tilde{\Upsilon}(\tau)$ and $\Omega(\tau)$ for $0 \leq \tau < t$ be introduced. Then the equation (21) can be rewritten as (22).

$$\tilde{\Upsilon}(\tau) = \hat{\theta}^T(t)\Omega(\tau) - \theta^T\Omega(\tau) - \varepsilon(\tau) = \tilde{\theta}^T(t)\Omega(\tau) - \varepsilon(\tau) \quad (22)$$

Now it is possible to obtain the adaptation law for $\theta$. For this purpose, only for the synthesis process, we will consider the case when $\varepsilon(\tau) = 0$, and introduce the optimization objective criterion (23) using the equation (22), which reflects the difference between the model $\hat{\Upsilon}(t)$ and the ideal function output $\Upsilon(t)$ [2, 15].

$$Q(\hat{\theta}) = \frac{1}{2}\int_0^t e^{-\lambda(t-\tau)}\tilde{\Upsilon}^T(\tau)\tilde{\Upsilon}(\tau)\,d\tau, \quad (23)$$

where $\lambda$ is an exponential forgetting factor [15].

The objective criterion (23) minimum is found when its gradient with respect to the adjustable parameters is equal to zero (24).

$$\nabla_{\hat{\theta}^T} Q^T\left(\hat{\theta}\right) = \int_0^t e^{-\lambda(t-\tau)} \Omega(\tau) \left[\Omega^T(\tau)\hat{\theta}(t) - \Upsilon^T(\tau)\right] d\tau = 0 \quad (24)$$

Considering (24) and using the property of the sum of integrals, we expand the brackets and move the term, containing the ideal value $\Upsilon(t)$, to the right side of the equality to obtain (25).

$$\int_0^t e^{-\lambda(t-\tau)} \Omega(\tau)\Omega^T(\tau)\hat{\theta}(t) d\tau = \int_0^t e^{-\lambda(t-\tau)} \Omega(\tau)\Upsilon^T(\tau) d\tau \quad (25)$$

Then the least-squares estimate of the ideal parameters can be written as (26).

$$\hat{\theta}(t) = \underbrace{\left[\int_0^t e^{-\lambda(t-\tau)} \Omega(\tau)\Omega^T(\tau) d\tau\right]^{-1}}_{\Gamma(t)} \int_0^t e^{-\lambda(t-\tau)} \Omega(\tau)\Upsilon^T(\tau) d\tau \quad (26)$$

Here $\Gamma(t)$ is an adaptation rate matrix.

The equation for $\Gamma^{-1}(t)$ derivative (27) can be found using the main theorem of calculus.

$$\frac{d\Gamma^{-1}}{dt} = \Omega(t)\Omega^T(t) - \lambda\int_0^t e^{-\lambda(t-\tau)} \Omega(\tau)\Omega^T(\tau) d\tau = \Omega(t)\Omega^T(t) - \lambda\Gamma^{-1}(t) \quad (27)$$

The equation for the $\Gamma(t)$ derivative (29) can be found using (28) and the above-introduced definitions of $\Gamma(t)$ and $\Gamma^{-1}(t)$.

$$\frac{dI}{dt} = \frac{d}{dt}\left[\Gamma(t)\Gamma^{-1}(t)\right] = \frac{d\Gamma(t)}{dt}\Gamma^{-1}(t) + \frac{d\Gamma^{-1}(t)}{dt}\Gamma(t) = 0 \quad (28)$$

$$\frac{d\Gamma(t)}{dt} = -\Gamma(t)\frac{d\Gamma^{-1}(t)}{dt}\Gamma(t) = \lambda\Gamma(t) - \Gamma(t)\Omega(t)\Omega^T(t)\Gamma(t) \quad (29)$$

The adaptation law (30) can be written if we differentiate (26) and apply the main theorem of calculus.

$$\frac{d\hat{\theta}(t)}{dt} = \frac{d\Gamma(t)}{dt}\int_0^t e^{-\lambda(t-\tau)}\Omega(\tau)\Upsilon^T(\tau)d\tau + \Gamma(t)\frac{d}{dt}\left[\int_0^t e^{-\lambda(t-\tau)}\Omega(\tau)\Upsilon^T(\tau)d\tau\right] =$$
$$= \left(\lambda - \Gamma(t)\Omega(t)\Omega^T(t)\right)\hat{\theta}(t) - \lambda\hat{\theta}(t) + \Gamma(t)\Omega(t)\Upsilon^T(t) = \quad (30)$$
$$= \Gamma(t)\Omega(t)\left[\Upsilon^T(t) - \Omega^T(t)\hat{\theta}(t)\right] = -\Gamma(t)\Omega(t)\tilde{\Upsilon}^T(t)$$

Thus, the complete adaptation loop (31) to estimate the ideal plant (3) parameters is described by two adaptation laws: 1) the one for $\hat{\theta}$ (30), 2) the one for $\dot{\Gamma}$ (29).

$$\dot{\hat{\theta}} = -\Gamma\Omega\tilde{\Upsilon}^T$$
$$\dot{\Gamma} = \lambda\Gamma - \Gamma\Omega\Omega^T\Gamma \quad (31)$$

We formulate the properties of the adaptation loop (31) in two theorems. The first one is about the parameter error $\tilde{\theta}$, and the second one is about the adaptation rate matrix $\Gamma$.

*Theorem 1.* Consider the adaptation loop (31) to estimate θ. The following properties hold for $\tilde{\theta}$:

1) if $\varepsilon(t) = 0$, then $\tilde{\theta}$ is bounded: $\tilde{\theta} \in L_2 \cap L_\infty$;

2) if $\varepsilon(t) = 0$, then the equations (31) provide exponential convergence of $\tilde{\theta}$ to zero with the rate, which is higher than $\kappa$ (its value is defined in the appendix and can be made arbitrarily high by the forgetting factor $\lambda$ value increase).

3) if $\varepsilon(t) \neq 0$, then $\tilde{\theta}$ is uniformly bounded by the set (32) with the ultimate bound $R$ and converges to it exponentially with the rate, which is not less than $0.5\kappa$.

$$B_R = \left\{\tilde{\theta}: \|\tilde{\theta}\| \leq R\right\}; \quad R = \frac{2\delta^2 \beta^{-1} \varepsilon_{max} \sqrt{n+m+1}}{\kappa \lambda_{max}(\Gamma^{-1})} \sqrt{\frac{\lambda_{max}(\Gamma^{-1})}{\lambda_{min}(\Gamma^{-1})}} \quad (32)$$

All points of the theorem 1 comply with the problem statement given in the respective section. The proof of the theorem 1 is given in the Appendix.

*Theorem 2.* The adaptation law (29) for $\Gamma$ provides the following properties:

1) if $t < T$, then the norm of the adaptation rate matrix $\Gamma$ is bounded in accordance with (33).

$$\|\Gamma\| \leq \sqrt{e^{2\lambda(T-t_1)} L(0)} \quad (33)$$

2) if $t \geq T$ and $t \to \infty$, then the equation (34) for the norm of the adaptation rate matrix $\Gamma$ holds true.

$$\lim_{t \to \infty} \|\Gamma\| = \frac{\lambda}{\lambda_{min}^2(\Omega(t))}; \quad (34)$$

**On one aspect of practical implementation**

The regressor and output signal derivatives are calculated with equation (15), which contains the exponentially damping function. So, when the experiment time $t > (3 \div 5)\beta^{-1}$, then the adaptation loop (31) no longer takes into account new data to identify the plant parameters. This makes the developed scheme ideally suitable to be applied for real plants, which function according to the step-like setpoint schedule $r$, which consists of a set of different constants. It is necessary to equal to zero the time variable in the filter formulas (15) at each moment when the setpoint $r$ is changed.

**Experimental results**

The efficiency of the proposed method has been shown by means of mathematical modeling of the identification process of the plant (35) parameters. Modeling was conducted in Matlab/Simulink on the basis of numerical integration by the Euler method. A constant step size of $\tau_s = 10^{-4}$ seconds was used in all experiments.

$$y(t) = \frac{4s+1}{s^2+s+4} u(t) \quad (35)$$

The filters parameters in (3) for this plant were chosen as follows – $\lambda_0 = 15$; $\lambda_1 = 45$. The adaptation loop parameters were the same (unless otherwise specified) in all experiments: $\Gamma_0 = I$; $\lambda = 1$; $\beta = 2$. The initial value of $\hat{\theta}$ was assumed to be zero. In each experiment, one of the adaptation loop (31) properties (defined in *Theorems* 1 and 2) was checked. In all experiments, a constant signal $u(t) = const = 100$ was used as the plant (35) input.

The first experiment was to check whether the developed adaptation loop provided exponential convergence of $\hat{\theta}$ values to the ideal ones of the plant (35) (initial conditions and noise were equal to zero). Fig.1 shows the comparison of the norms of the parameter error $\tilde{\theta}$, obtained with the help of (31) and the classical gradient scheme (7), for which the constant adaptation rate was used: $\Gamma = \Gamma_0 = I$.

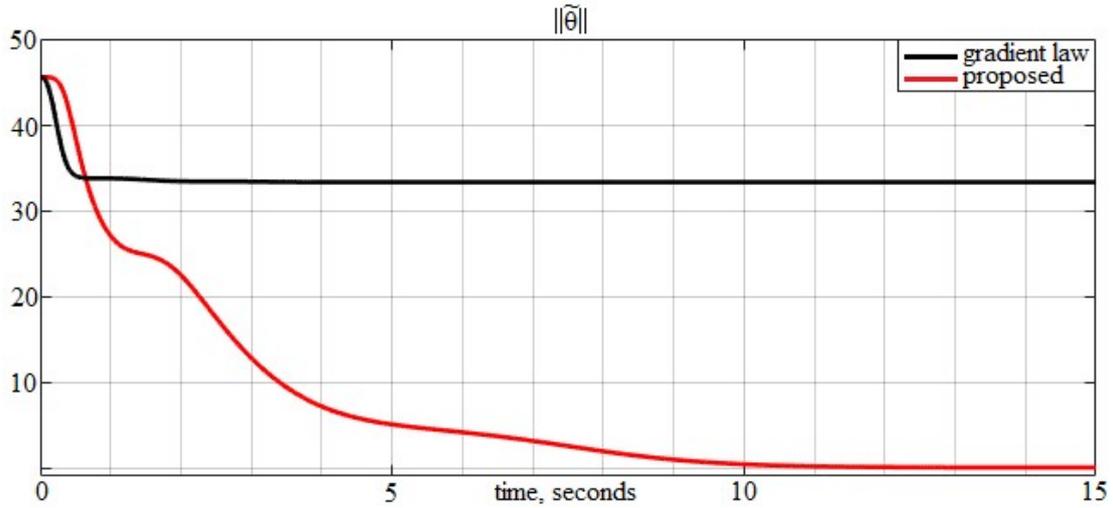
Fig.1. Norm of parameter error $\tilde{\theta}$ of developed adaptation loop and gradient scheme

As it follows from Fig.1, the developed scheme, unlike the gradient one, provided exponential convergence of the norm of the parameter error $\tilde{\theta}$ to zero when the control signal was constant $u(t) = const = 100$.

Fig.2 shows the norm of the adaptation rate matrix $\Gamma$ (red curve), its upper bound (33) + positive invariant (34) (black curve).

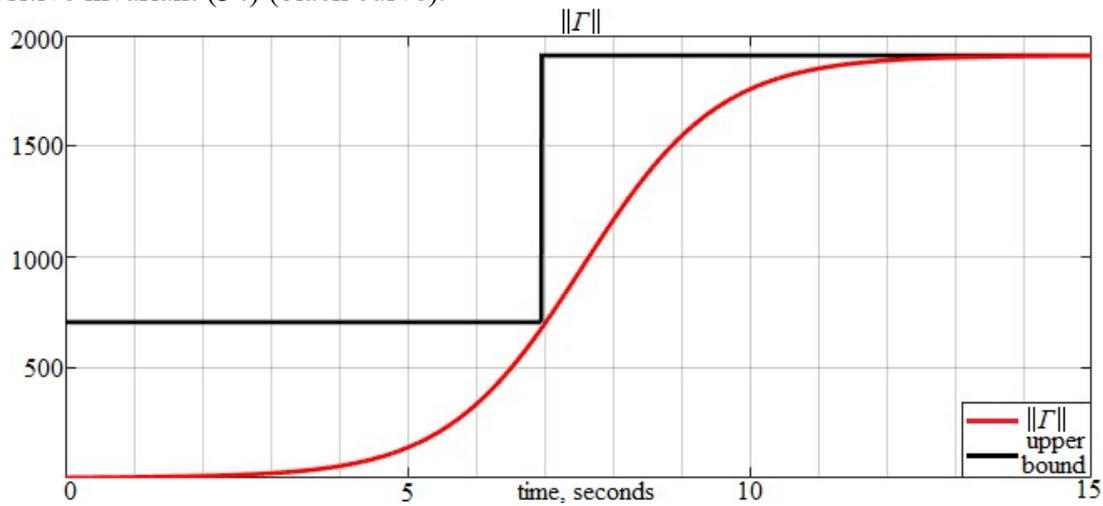
Fig.2. Norm of adaptation rate matrix $\Gamma$

As it follows from Fig. 2, the equations (33) and (34) held true.

Fig.3 shows a comparison of the norms of the parameter error $\tilde{\theta}$ obtained using the developed adaptation loop with different values of the forgetting factor $\lambda$.

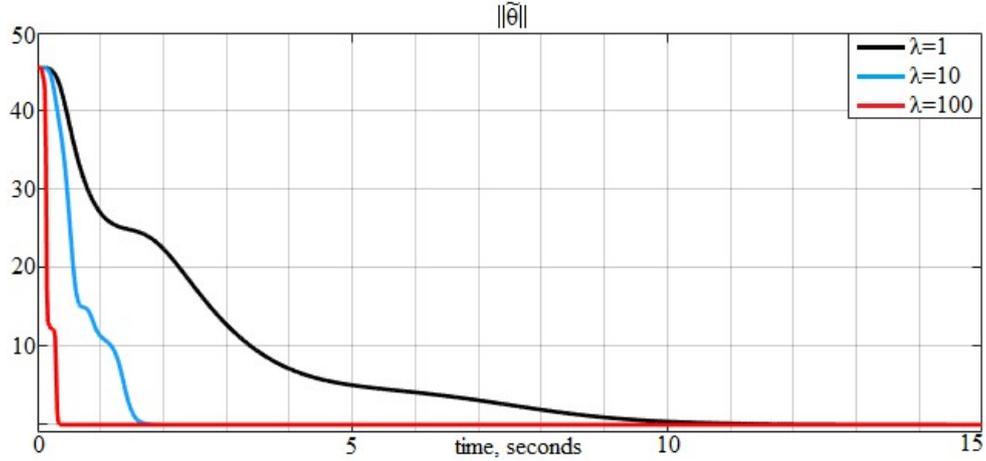

Fig.3. Norm of parameter error $\tilde{\theta}$ obtained using developed adaptation loop with different values of $\lambda$

Fig. 3 shows that the convergence rate of the plant (35) parameters estimation process can indeed be improved by increasing the forgetting factor $\lambda$ value. This corresponds to the conclusion made in *Remark* 5 (in the Appendix).

Fig.4A and Fig.4B present a comparison of the norms of the parameter error $\tilde{\theta}$ and minimum eigenvalues $\lambda_{min}^2(\Omega(t))$ obtained using different values of the memory factor $\beta$.

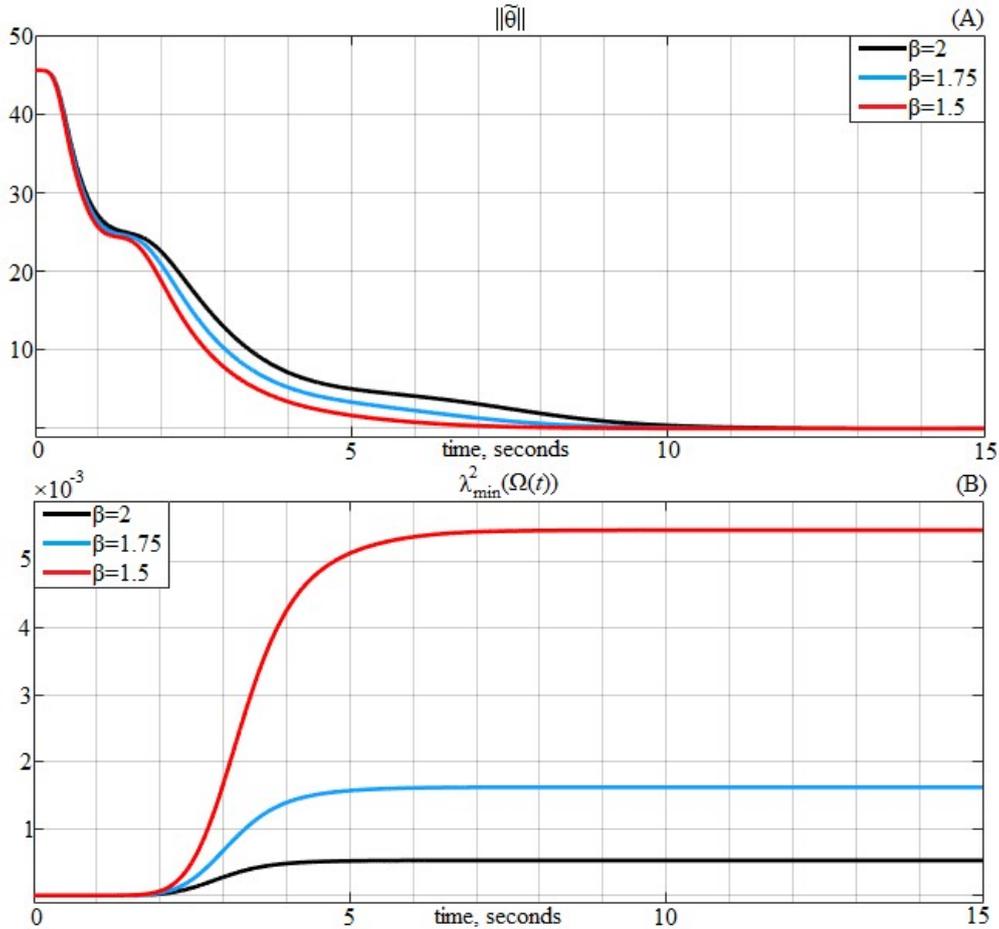

Fig.4. Comparison of norms of parameter error $\tilde{\theta}$ and minimum eigenvalues $\lambda_{min}^2(\Omega(t))$ obtained using different values of memory factor $\beta$

As it follows from Fig.4B, if the memory factor β was decreased, then the minimum eigenvalue $\lambda^2_{min}(\Omega(t))$ became higher. This, in its turn, led to the improvement of the convergence rate κ (see Fig.4A and (A.9)).

The second experiment was to check the robustness of the parameter estimation with respect to ε(t), which can be caused by the initial conditions and/or the disturbances. In this experiment, the function ε(t) was caused by the measurement noise. It was simulated as white noise with parameters: *Noise Power* = 100; *Seed* = 23341·10³; *Sample Time* $\tau_s = 10^{-4}$ seconds. Fig.5 shows a comparison of the norms of the parameter error $\tilde{\theta}$ obtained using the developed adaptation loop and the integral PI law proposed in [13], for which the constant adaptation rate was used, which value coincided with the initial rate for the developed loop ($\Gamma = \Gamma_0 = I$).

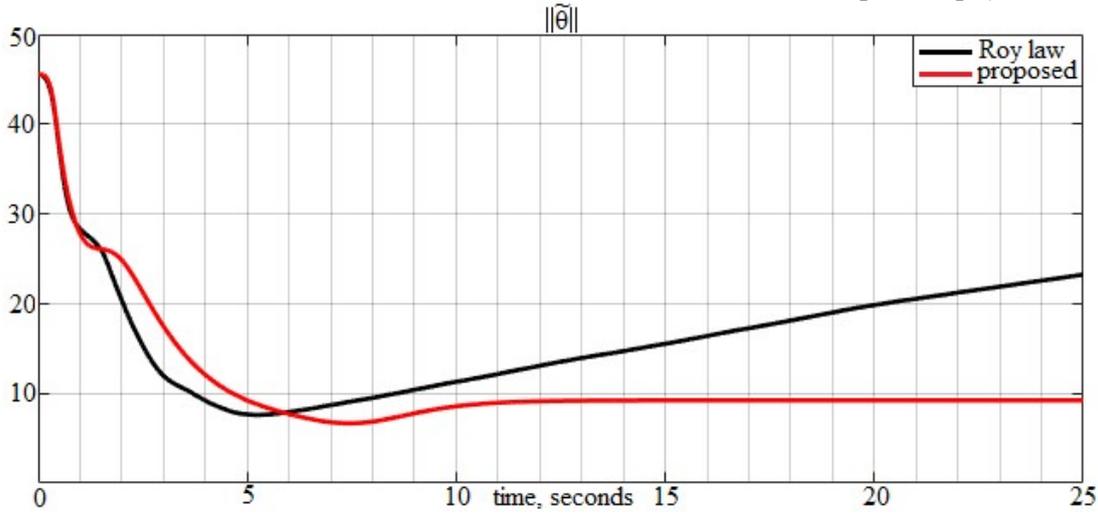

Fig.5. Comparison of norms of parameter error $\tilde{\theta}$ obtained using developed adaptation loop and integral PI law [13] in the case ε(t) ≠ 0

As it follows from Fig.5, the norm of the parameter error obtained using the integral PI law drifted. As for the developed method, the considered norm was bounded and converged to the neighborhood of zero.

Fig.6 shows a comparison of the norms of the parameter error $\tilde{\theta}$ obtained using the developed adaptation loop with different values of the memory factor β.

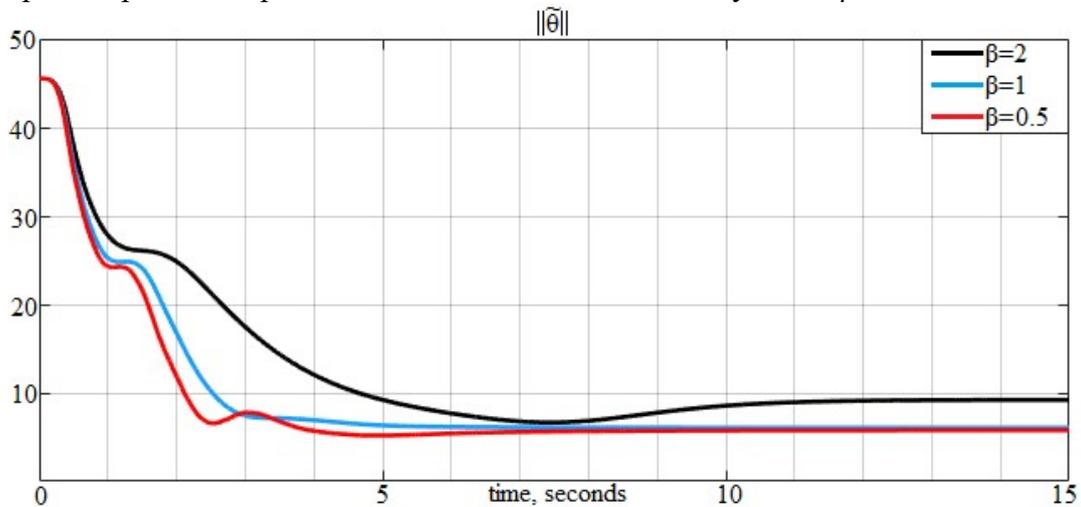

Рис.6. Norm of parameter error $\tilde{\theta}$ obtained using developed adaptation loop with different values of memory factor β in the case ε(t) ≠ 0

As it follows from Fig. 6, the neighborhood of zero, to which the norm of the parameter error converges, can be made small, but not zero, by reduction of the memory factor β value. "Not zero" because the too low value of the memory factor β, as it can be seen from (20), leads, in the case of the measurement noise, to $\varepsilon(t) \notin L_\infty$. This means that the attractive set (32) degenerates.

Thus, the conducted experiments have fully confirmed the properties of the adaptation loop (31) defined in *Theorems* 1 and 2.

## Conclusion

In this research, the adaptation loop was proposed to identify the LTI plant parameters, which did not require to meet the persistent excitation condition to provide exponential convergence of the parameter error to zero. The PE condition was relaxed to the initial excitation (IE) one. The robustness properties of the developed adaptation loop to the influence of initial conditions and bounded disturbances were proved. The convergence rate of the parameter error can be made arbitrarily high.

The further research scope is to apply the developed adaptation loop to the problem of the direct adaptive control with the reference model.

**APPENDIX**

**Theorem 1 proof.**

To prove the first and second parts of *Theorem* 1, the equation (21) is substituted into the equation (30) taking into account $\varepsilon(t) = 0$. Then, if $\theta$ = const, the equation (A.1) is obtained.

$$\dot{\tilde{\theta}} = -\Gamma \Omega \Omega^T \tilde{\theta} \tag{A.1}$$

The Lyapunov function candidate is chosen as a quadratic function (A.2).

$$V = \tilde{\theta}^T \Gamma^{-1} \tilde{\theta}$$

$$\lambda_{min}(\Gamma^{-1}) \|\tilde{\theta}\|^2 \leq V \leq \lambda_{max}(\Gamma^{-1}) \|\tilde{\theta}\|^2 \tag{A.2}$$

Considering (A.1) and (27), the derivative of (A.2) is (A.3).

$$\dot{V} = 2\tilde{\theta}^T \Gamma^{-1} \dot{\tilde{\theta}} + \tilde{\theta}^T \dot{\Gamma}^{-1} \tilde{\theta} =$$

$$= -2\tilde{\theta}^T \Gamma^{-1} [\Gamma \Omega \Omega^T \tilde{\theta}] + \tilde{\theta}^T [\Omega \Omega^T - \lambda \Gamma^{-1}] \tilde{\theta} = -\tilde{\theta}^T \Omega \Omega^T \tilde{\theta} - \lambda \tilde{\theta}^T \Gamma^{-1} \tilde{\theta} = \tag{A.3}$$

$$= -\tilde{\Upsilon}\tilde{\Upsilon}^T - \lambda \tilde{\theta}^T \Gamma^{-1} \tilde{\theta} \leq -\|\tilde{\Upsilon}\|^2 - \lambda \cdot \lambda_{min}(\Gamma^{-1}) \|\tilde{\theta}\|^2$$

*Proposition 1.* The derivative (A.3) of the positive definite quadratic form (A.2) is a negative semi-definite function, so parameter error $\tilde{\theta} \in L_\infty$ and error $\tilde{\Upsilon} \in L_\infty$, and equation (A.2) is the Lyapunov function for (A.1). At the same time, the function (A.2) has the finite limit (A.4) when $t \to \infty$, so $\tilde{\theta} \in L_2 \cap L_\infty$.

$$V(\tilde{\theta}(t \to \infty)) = V(\tilde{\theta}(0)) + \int_{t_0}^{\infty} \dot{V} \, dt = V(\tilde{\theta}(0)) - \int_{t_0}^{\infty} \left[ \tilde{\Upsilon}\tilde{\Upsilon}^T + \lambda(\tilde{\theta}^T \Gamma^{-1} \tilde{\theta}) \right] dt$$

$$\Rightarrow \int_{t_0}^{\infty} \left[ \|\tilde{\Upsilon}\|^2 + \lambda \cdot \lambda_{min}(\Gamma^{-1}) \|\tilde{\theta}\|^2 \right] dt = V(\tilde{\theta}(0)) - V(\tilde{\theta}(t \to \infty)) < \infty \tag{A.4}$$

Hence, the first part of the theorem is proved.

In order to prove the second part of *Theorem* 1, the equation (A.3) is rewritten as (A.5).

$$\dot{V} = -\tilde{\theta}^T \Omega \Omega^T \tilde{\theta} - \lambda \tilde{\theta}^T \Gamma^{-1} \tilde{\theta} \quad (A.5)$$

As the new regressor $\Omega(t)$ is a positive semi-definite function, and, taking into account the Definition 3, the initial regressor $\omega(t)$ is excited into the initial period of time (11), then for the new regressor $\Omega(t)$ there exists $T > 0$ and $\lambda_{min}(\Omega(T)) > 0$, so that, when $t \geq T$, the lower bound (A.6) holds true.

$$\Omega(t) = \int_0^t e^{-\beta\tau} \omega(\tau) \omega^T(\tau) d\tau \geq \underbrace{\int_0^T e^{-\beta\tau} \omega(\tau) \omega^T(\tau) d\tau}_{\Omega(T)} \geq \lambda_{min}(\Omega(T)) I \quad (A.6)$$

Then the lower bound (A.7) holds true for $\Omega(t)\Omega^T(t)$ when $t \geq T$. It follows from (A.7), that the initial excitation condition (A.8) is met for the new regressor $\Omega(t)$.

$$\Omega(t)\Omega^T(t) \geq \lambda_{min}^2(\Omega(T)) I \quad (A.7)$$

$$\int_0^T \Omega(\tau)\Omega^T(\tau) d\tau \geq \lambda_{min}^2(\Omega(T)) I \quad (A.8)$$

*Remark 4*: It follows from (A.6)-(A.8) that, if $\omega \in IE$, then also $\Omega(t) \in IE$. According to *Remark* 1, the condition of the initial excitation can be checked by the calculation of the new regressor $\Omega(t)$ determinant.

Thus, taking into account equation (A.7), the upper bound of the derivative (A.5) becomes (A.9) at $t \geq T$.

$$\dot{V} = -\tilde{\theta}^T \Omega \Omega^T \tilde{\theta} - \lambda \tilde{\theta}^T \Gamma^{-1} \tilde{\theta} \leq -\lambda_{min}^2(\Omega(T)) \|\tilde{\theta}\|^2 - \lambda \cdot \lambda_{min}(\Gamma^{-1}) \|\tilde{\theta}\|^2 \leq$$

$$\leq -\left[ \frac{\lambda_{min}^2(\Omega(T)) + \lambda \cdot \lambda_{min}(\Gamma^{-1})}{\lambda_{max}(\Gamma^{-1})} \right] \lambda_{max}(\Gamma^{-1}) \|\tilde{\theta}\|^2 \leq -\kappa V, \quad (A.9)$$

$$\kappa = \left[ \frac{\lambda_{min}^2(\Omega(T)) + \lambda \cdot \lambda_{min}(\Gamma^{-1})}{\lambda_{max}(\Gamma^{-1})} \right]$$

The lower bound of the Lyapunov function is substituted into (A.9). Then the whole equation is integrated. The equation (A.10) is obtained as a result.

$$\|\tilde{\theta}\| \leq \sqrt{\lambda_{min}^{-1}(\Gamma^{-1}) e^{-\kappa \cdot t} V(0)} \quad (A.10)$$

It follows from (A.10) that, when $\varepsilon(t)=0$, the error $\tilde{\theta}$ converges to zero exponentially and faster than $\kappa$. The second part of *Theorem* 1 is proved.

*Remark 5*: As it follows from the equation for $\kappa$, its value can be made arbitrary high as a result of the forgetting factor $\lambda$ value increase.

To prove the third part of the theorem, the equation (21) is substituted into the equation (30) taking into account $\varepsilon(t) \neq 0$. Then the equation (A.11) is obtained.

$$\dot{\tilde{\theta}} = -\Gamma \Omega \Omega^T \tilde{\theta} + \Gamma \Omega \varepsilon^T \quad (A.11)$$

Considering (A.11) and (27), the derivative of (A.2) is (A.12).

$$\dot{V} = 2\tilde{\theta}^T \Gamma^{-1} \left[ -\Gamma \Omega \Omega^T \tilde{\theta} + \Gamma \Omega \varepsilon^T \right] + \tilde{\theta}^T \left[ \Omega \Omega^T - \lambda \Gamma^{-1} \right] \tilde{\theta} =$$
$$= -\tilde{\theta}^T \Omega \Omega^T \tilde{\theta} - \lambda \tilde{\theta}^T \Gamma^{-1} \tilde{\theta} + 2\tilde{\theta}^T \Omega \varepsilon^T \quad (A.12)$$

The upper bounds of the regressor $\Omega(t)$ (19) and disturbance $\varepsilon(t)$ (20) are used to obtain the upper bound of the derivative (A.12) as (A.13) when $t \geq T$.

$$\dot{V} \leq -\lambda_{min}^2\left(\Omega(T)\right)\|\tilde{\theta}\|^2 - \lambda \cdot \lambda_{min}\left(\Gamma^{-1}\right)\|\tilde{\theta}\|^2 + 2\delta^2\beta^{-1}\varepsilon_{max}\sqrt{n+m+1}\|\tilde{\theta}\| \leq$$
$$\leq -\kappa\lambda_{max}\left(\Gamma^{-1}\right)\|\tilde{\theta}\|^2 + 2\delta^2\beta^{-1}\varepsilon_{max}\sqrt{n+m+1}\|\tilde{\theta}\| \tag{A.13}$$

Hence, the derivative (A.13) is negative outside the compact set (A.14).

$$B_r = \left\{\tilde{\theta}: \|\tilde{\theta}\| \leq \frac{2\delta^2\beta^{-1}\varepsilon_{max}\sqrt{n+m+1}}{\kappa\lambda_{max}\left(\Gamma^{-1}\right)} = r\right\} \tag{A.14}$$

Given the boundedness of the chosen Lyapunov function candidate (A.2), the definition (A.15) of the minimum and maximum level lines of the function (A.2) is introduced.

$$\begin{aligned}C_{min} &= \left\{\tilde{\theta} \in \partial B_r: V = c_{min}\right\} \\ C_{max} &= \left\{\tilde{\theta} \in R^{n+m+1}: V = c_{max}\right\}\end{aligned} \tag{A.15}$$

Here $\partial B_r$ is the boundary of the set (A.14). The minimum value of the Lyapunov function is reached at the boundary $\partial B_r$ of $B_r$, because the derivative (A.13) is positive inside $B_r$ and negative outside it. Using the definition of the level lines of the Lyapunov function, the annulus set (A.16) is introduced.

$$\Lambda = \left\{\tilde{\theta}: c_{min} \leq V \leq c_{max}\right\} \tag{A.16}$$

*Proposition 2.* As the derivative (A.13) of (A.2) is negative inside the annulus and positive inside $C_{min}$, so: 1) if the trajectories of $\tilde{\theta}$ start in (A.16), then they will enter the $C_{min}$ in finite time; 2) if the trajectories of $\tilde{\theta}$ start in $C_{min}$, they will not leave it. As the derivative (A.13) is positive inside $C_{min}$, then the boundedness of the trajectories is to be shown inside it.

According to the boundedness of (A.2), the inequality (A.17) holds true for $\forall \tilde{\theta} \in B_r$: the quadratic form (A.2) reaches its maximum value $\forall \tilde{\theta} \in B_r$ when $\tilde{\theta}$ has its maximum value from $B_r$. In its turn, that is true when $\tilde{\theta} \in \partial B_r$.

$$V \leq \lambda_{max}\left(\Gamma^{-1}\right)\|r\|^2 \tag{A.17}$$

Then, according to the definition of $C_{min}$, we obtain (A.18).

$$c_{min} = \lambda_{max}\left(\Gamma^{-1}\right)\|r\|^2 \tag{A.18}$$

As (A.2) is bounded, the inequality (A.19) holds true for $\forall \tilde{\theta} \in C_{min}$.

$$\lambda_{min}\left(\Gamma^{-1}\right)\|\tilde{\theta}\|^2 \leq V \leq \lambda_{max}\left(\Gamma^{-1}\right)\|r\|^2 \tag{A.19}$$

It follows from inequalities (A.19) that the trajectories of $\tilde{\theta}$ are bounded by the set (32), and their ultimate bound $R$ is (A.20).

$$R = \|r\|\sqrt{\frac{\lambda_{max}\left(\Gamma^{-1}\right)}{\lambda_{min}\left(\Gamma^{-1}\right)}} \tag{A.20}$$

To estimate the rate of convergence of $\tilde{\theta}$ to the set (32), we complete the square of the right side of the inequality (A.13) to obtain (A.21).

$$-\kappa\lambda_{max}\left(\Gamma^{-1}\right)\|\tilde{\theta}\|^2 + 2\delta^2\beta^{-1}\varepsilon_{max}\sqrt{n+m+1}\|\tilde{\theta}\| \le$$

$$\le \frac{1}{2}\left[-\kappa\lambda_{max}\left(\Gamma^{-1}\right)\|\tilde{\theta}\|^2 - \left(\sqrt{\kappa\lambda_{max}\left(\Gamma^{-1}\right)}\|\tilde{\theta}\| - \frac{2\delta^2\beta^{-1}\varepsilon_{max}\sqrt{n+m+1}}{\sqrt{\kappa\lambda_{max}\left(\Gamma^{-1}\right)}}\right)^2 + \right. \quad (A.21)$$

$$\left. + \frac{4\delta^4\beta^{-2}\varepsilon_{max}^2(n+m+1)}{\kappa\lambda_{max}\left(\Gamma^{-1}\right)}\right] \le \frac{-\kappa\lambda_{max}\left(\Gamma^{-1}\right)\|\tilde{\theta}\|^2}{2} + \frac{2\delta^4\beta^{-2}\varepsilon_{max}^2(n+m+1)}{\kappa\lambda_{max}\left(\Gamma^{-1}\right)}$$

Taking into consideration (A.21), the upper bound of the derivative (A.12) is written as (A.22).

$$\dot{V} \le \frac{-\kappa\lambda_{max}\left(\Gamma^{-1}\right)\|\tilde{\theta}\|^2}{2} + \frac{2\delta^4\beta^{-2}\varepsilon_{max}^2(n+m+1)}{\kappa\lambda_{max}\left(\Gamma^{-1}\right)} \quad (A.22)$$

Considering the definition of the Lyapunov function candidate (A.2), the equation (A.22) is integrated with respect to time to obtain (A.23).

$$\lambda_{min}\left(\Gamma^{-1}\right)\|\tilde{\theta}\|^2 \le V \le e^{-0.5\kappa t}V(0) + \frac{4\delta^4\beta^{-2}\varepsilon_{max}^2(n+m+1)}{\kappa^2\lambda_{max}\left(\Gamma^{-1}\right)} \quad (A.23)$$

It follows from (A.23), and the left bound of the Lyapunov function (A.2), and the fact that the first term of the right side of (A.23) converges to zero when $t\to\infty$, that the parameter error converges to the set (32) exponentially and faster than $0{,}5\kappa$. So the third part of *Theorem 1* is proved.

**Theorem 2 proof**

To prove the second theorem, the Lyapunov function candidate (A.24) for (29) is introduced.

$$L = tr\left(\Gamma^T\Gamma\right) = \|\Gamma\|^2 \quad (A.24)$$

Considering (29), the derivative of (A.24) is written as (A.25).

$$\dot{L} = tr\left(2\Gamma^T\dot{\Gamma}\right) = tr\left(2\Gamma^T\left[\lambda\Gamma - \Gamma\Omega\Omega^T\Gamma\right]\right) = 2\lambda L - 2L\lambda_{min}\left(\Omega\Omega^T\right)\|\Gamma\| \quad (A.25)$$

Let the case be considered when $t < T$ to prove the first part of the theorem. Then $\lambda^2_{min}(\Omega(t)) \approx 0$, because the IE condition (A.8) has not been met yet. So the upper bound of the derivative (A.25) is (A.26).

$$\dot{L} \le 2\lambda L \quad (A.26)$$

The equation (A.26) is integrated with respect to time between 0 and $T - t_1$, where $t_1 > 0$ and $T \gg t_1$, to obtain (A.27).

$$L = \|\Gamma\|^2 \le e^{2\lambda(T-t_1)}L(0) \quad (A.27)$$

Using (A.27), the estimation (33) of the norm of the adaptation rate matrix is obtained.

Then let the case be considered when $t \ge T$ to prove the second part of the theorem 2. Then $\lambda^2_{min}(\Omega(t)) > \lambda^2_{min}(\Omega(T)) > 0$, because the IE condition (A.8) has already been met. Taking into account (A.6) and (A.7), in this case, the derivative (A.25) is written as (A.28).

$$\dot{L} = 2\lambda L - 2L\lambda^2_{min}\left(\Omega(t)\right)\|\Gamma\| = 2\lambda L - 2L\lambda^2_{min}\left(\Omega(t)\right)\sqrt{L} \le 0 \quad (A.28)$$

It follows from (A.28) that the derivative of (A.24) is negative outside the compact set (A.29).

$$C_d = \left\{ \Gamma \in R^{(n+m+1)\times(n+m+1)} \colon \|\Gamma\| \leq \frac{\lambda}{\lambda_{min}^2(\Omega(t))} = d \right\} \tag{A.29}$$

As the derivative of the quadratic form $L$ is negative outside the $C_d$ set and positive inside it, then, according to the LaSalle theorem [1, 2], the $\partial C_d$ set (A.30), which defines the boundary of the $C_d$ set, is a positive invariant for the trajectories of equation (29).

$$\partial C_d = \left\{ \Gamma \in C_d \colon \|\Gamma\| = d \right\} \tag{A.30}$$

The analytical proof that the trajectories of equation (29) at $t \geq T$ and $t \to \infty$ really converge to $d$ is shown below. A fractional decomposition is applied to the upper bound (A.28) to obtain (A.31). Then it is integrated with respect to time and solved for $L$ (A.32).

$$\frac{dL}{2L\left[-\lambda + \lambda_{min}^2(\Omega(t))\sqrt{L}\right]} = -dt$$

$$\left( \frac{0{,}5\lambda^{-1}\lambda_{min}^2(\Omega(t))L^{\frac{-1}{2}}}{-\lambda + \lambda_{min}^2(\Omega(t))\sqrt{L}} - \frac{\lambda^{-1}}{2L} \right) dL = -dt \tag{A.31}$$

$$L = \frac{\lambda^2 e^{2\lambda t}}{\left( \lambda_{min}^2(\Omega(t))e^{\lambda t} + e^{\lambda/2} \right)^2} \tag{A.32}$$

Having used L'Hospital rule twice, it is shown that when $t \geq T$ and $t \to \infty$, then the quadratic form $L$ converges to $d^2$ – (A.33).

$$\lim_{t \to \infty} L = \lim_{t \to \infty} \left( \frac{\lambda^2 e^{2\lambda t}}{\left( \lambda_{min}^2(\Omega(t))e^{\lambda t} + e^{\lambda/2} \right)^2} \right) = \frac{\lambda^2}{\lambda_{min}^4(\Omega(t))} = d^2 \tag{A.33}$$

Then, using the definition of the quadratic form $L$, equality (34) holds true. So the second part of *Theorem* 2 is proved.

Research was financially supported by Russian Foundation for Basic Research (Grant No 18-47-310003-r-a).